\documentclass[aps,twocolumn,groupedaddress,showpacs]{revtex4}
\usepackage[dvips]{graphicx}
\usepackage{amsmath}
\usepackage{amssymb}
\begin{document}
\bibliographystyle{prsty}
\title{Perturbations in a regular bouncing Universe}
\author{T.~J. Battefeld $^{1)}$\footnote{battefeld@physics.brown.edu}
and  G. Geshnizjani $^{2)}$\footnote{ghazal@physics.wisc.edu}}
\affiliation{1) Physics Department, Brown University,
  Providence RI 02912 USA,}
\affiliation{2) University of Wisconsin, Madison WI 53706 USA.}
\date{\today}
\preprint{hep-th/0503160}
\preprint{BROWN-HET-1443}
\begin{abstract}
We consider a simple toy model of a regular bouncing universe. The bounce is caused by an extra time-like dimension, which leads to a sign flip of the $\rho^2$ term in the effective four dimensional Randall Sundrum-like description. We find a wide class of possible bounces: big bang avoiding ones for regular matter content, and big rip avoiding ones for phantom matter.

Focusing on radiation as the matter content, we discuss the
evolution of scalar, vector and tensor perturbations. We compute a
spectral index of $n_s=-1$ for scalar perturbations and a deep blue index for tensor perturbations after
invoking vacuum initial conditions, ruling out such a model as a
realistic one. We also find that the spectrum (evaluated at Hubble
crossing) is sensitive to the bounce. We conclude that it is
challenging, but not impossible, for cyclic/ekpyrotic models to
succeed, if one can find a regularized version.
\end{abstract}
\maketitle

\section{Introduction}
Recently, bouncing models of the universe were revived in the framework of string cosmology \cite{Lidsey:1999mc}, since it is possible to generate a scale invariant spectrum of density fluctuations in the pre-bounce phase. Specific realizations are e.g. cyclic/ekpyrotic models of the universe \cite{Khoury:2001wf,Khoury:2001bz,Steinhardt:2001st,Gratton:2003pe,Tolley:2003nx} or the pre-big bang scenario \cite{Lidsey:1999mc,Gasperini:1992em,Gasperini:2002bn}. The main problem for such models to succeed lies in the fact that the bounce itself is often singular in the models studied so far \cite{Turok:2004gb,Battefeld:2004mn}. This makes matching conditions a necessity \cite{Khoury:2001zk,Brandenberger:2001bs,Durrer:2002jn,Martin:2001ue} which are, unfortunately, rather ambiguous. In addition, fluctuations often become non perturbative near a singularity \cite{Goheer:2004gk}.
In the few toy models of regular bouncing cosmologies known in the literature, the actual bounce has a strong impact on the evolution of perturbations \cite{Peter:2004um,Martin:2003sf,Allen:2004vz,Gasperini:2003pb,Gasperini:2004ss,Bozza:2005wn,Bozza:2005xs,Bozza:2005qg,Peter:2002cn,Finelli:2003mc,Cartier:2004zn}. What is more, the models are often technically challenging at the perturbative level and unfortunately inconsistent methods were proposed in the literature \footnote{The splitting method used in v2 (on the arxiv) of this article is indeed inconsistent, as we showed in detail in \cite{Geshnizjani:2005hc}. As a consequence, the scalar perturbation section of the present article was considerably revised in v3.}. Consequently, bouncing models have not been taken seriously as alternatives to the inflationary paradigm \cite{Linde:2002ws}.

In this article we will construct a simple regular toy model, for which one can compute analytically how perturbations evolve through the bounce. We find a red spectral index for scalar perturbations that is ruled out by observations \cite{Spergel:2003cb}. Furthermore, the detailed mechanism providing the bounce has a strong impact on the spectrum.

In order to generate a bounce within pure four dimensional general relativity, one has to violate various energy conditions. The usual mechanism in the literature involves the addition of an extra exotic matter field that becomes dominant during the bounce \cite{Allen:2004vz,Bozza:2005wn,Bozza:2005xs,Peter:2002cn,Peter:2003,Finelli:2003mc} (see also \cite{Cartier:2004zn} for an $\alpha^\prime$ regularized bounce, \cite{Brandenberger:1993ef,Tsujikawa:2002qc} for a non-singular cosmology achieved by higher derivative modification of general relativity, or \cite{Gasperini:2003pb,Gasperini:2004ss} for a bounce induced by a general-covariant, $T$-duality-invariant, non-local dilaton potential). The presence of a second matter field, and perturbations in it, is one physical origin for the sensitivity to the type of bounce one is considering.

We will not add an additional matter field, but use an ingredient motivated by string theory: extra dimensions. If we assume one additional dimension, the effective four dimensional Friedmann equation will show corrections proportional to $\rho^2$ at high energy densities \footnote{There are also corrections due to the projected Weyl tensor, but we will ignore these in the following.}. If the extra dimension is space-like, we have the usual Randall Sundrum setup \cite{Randall:1999ee,Brax:2004xh}. In that case, $\rho$ and $\rho^2$ have the same sign, and no bounce will occur.

However, if we assume the additional dimension to be time-like,
this sign will flip and thus it will cause a bounce when $\rho^2$
becomes dominant \cite{Shtanov:2002mb,Brown:2004cs} \footnote{See also \cite{Piao:2004hr} where this mechanism is employed to study a universe dominated by a massive scalar field.}. We are aware
of complexities associated with an extra time-like dimension; for
instance, existence of tachyonic modes and possible violation of
causality, or possible appearance of negative norm states (see
\cite{Dvali:1999,Matsuda,Chaichian,Bars:2001xv} for more
discussion regarding these issues).
 We will not touch on those issues, but take the modification of the Friedmann equations as a simple toy model
 that could also arise by other means.
The outline of the article is as follows: first, we work out the
background solution based on the above idea in section
\ref{secbgr}. We find a whole class of simple, analytically known,
bouncing cosmologies, either big bang avoiding ones, for usual
matter content, or big rip avoiding ones, for phantom matter. We
will then focus on one specific big bang avoiding bounce by
considering radiation as the matter content -- this seems to be
the most conservative choice to us.

In Section \ref{secpert} we discuss scalar perturbations and compute
analytically the spectrum of fluctuations in the post bounce era,
for Bunch-Davis vacuum initial condition in the pre-bounce era. We
find a red spectrum with
 a spectral index of $n_s=-1$, thus ruling out this model as a realistic one.
 We also find that the background cosmology around the bounce dictates the deformation of the spectrum
 through the transition. In Section \ref{interpret}, we conclude that it is challenging for the bounce
in Cyclic/Ekpyrotic models of the universe to leave the spectrum
unaffected. The only way to check if this is the case is by finding
a regularized version of the proposed scenarios and following the
perturbations explicitly through the bounce.

In Section \ref{secvector}, we discuss vector perturbations and find
that they remain perturbative during the bounce. Finally in Section
\ref{sectensor}, we derive analytically the evolution of
gravitational waves as they pass through the bounce. We find a blue
spectrum of gravitational waves in the post bounce era on large
scales, and an amplitude that depends on the details of the bounce.

\section{Background \label{secbgr}}
We use a metric with negative signature, scale factor $a(t)$, cosmic time $t$ and,
for simplicity, we work with a flat universe, so that
\begin{eqnarray}
d\,s^2&=&dt ^2 -a^2\delta _{ij}d\,x^id\,x^j\,.
\end{eqnarray}
Considering a modified version of the Randall Sundrum (RS) scenario \cite{Randall:1999ee,Brax:2004xh},
originating e.g. by having an extra
time-like
dimension \cite{Shtanov:2002mb,Brown:2004cs} we have the modified Einstein equations
\begin{eqnarray}
G_{\mu\nu}=-\kappa^2T_{\mu\nu}+\tilde{\kappa}^4S_{\mu\nu}\,, \label{einstein}
\end{eqnarray}
where $\kappa^{2}=8\pi /M_p^2$, $\tilde{\kappa}^2=8\pi/\tilde{M}_p^3$ and $\tilde{M}_p$ is the fundamental
5-dimensional Planck mass. For simplicity and since we are primarily interested in the high energy modifications due to $S_{\mu\nu}$, we fine tuned the model such that the effective four dimensional cosmological constant vanishes and we neglected the projected Weyl
tensor.

$S_{\mu\nu}$ is quadratic in the energy momentum tensor $T_{\mu\nu}$ and is given by
\cite{Shiromizu:1999wj,Maartens:2000fg}
\begin{eqnarray}
\nonumber S_{\mu\nu}&=&\frac{1}{12}T^{\alpha}_{\,\,\alpha}T_{\mu\nu}-\frac{1}{4}T_{\mu\alpha}T^\alpha _{\,\,\nu}\\
&&+\frac{1}{24}g_{\mu\nu}\left[3T_{\alpha\beta}T^{\alpha\beta}-T^\alpha _{\,\,\alpha}T^{\beta}_{\,\,\beta}\right]
\,.     \label{smunu}
\end{eqnarray}
We have a different sign in front of $S_{\mu\nu}$ in (\ref{einstein}) compared
to the usual RS setup \cite{Brax:2004xh}. This modification
will yield a class of non singular bounces we shall examine below.
If we consider an ideal fluid with
\begin{eqnarray}
( T^{\mu}_{\,\,\nu})&=&
\left(
\begin{array}{cc}
-\rho&0\\
0&p \delta ^i_j
\end{array}
\right)\,,
\end{eqnarray}
the quadratic term becomes
\begin{eqnarray}
( S^{\mu}_{\,\,\nu})&=&
\left(
\begin{array}{cc}
-\frac{1}{12}\rho ^2&0\\
0&\frac{1}{12}\rho(\rho +2p)\delta ^i_j
\end{array}
\right)\,.
\end{eqnarray}
We may rephrase the quadratic corrections in terms of a second ideal fluid
with
\begin{eqnarray}
\rho _{-}&:=&\frac{\rho ^2}{2\lambda}\,,\\
p_{-}&:=&\frac{\rho}{2\lambda}\left(2p+\rho\right)\,,
\end{eqnarray}
where $\lambda=6\kappa^2/\tilde{\kappa}^4$. The "two fluids" are of course related and represent only one degree of freedom. If we have an equation of state between $\rho_{+}:=\rho$ and
 $p_{+}:=p$, that is
$p_+=w\rho_+$, we get $p_-=w_-\rho_-$ with $w_-=2w+1$. It will become useful
to write $w=(n-3)/3$ so that
$w_-=2n/3-1$. As a consequence of the equations of state we get
\begin{eqnarray}
\rho_+(t)=ra(t)^{-n}\,,\label{rho+}\\
\rho_-(t)=\frac{r^2}{2\lambda}a(t)^{-2n}\,.
\end{eqnarray}
The Friedmann equation reads now
\begin{eqnarray}
H^2=\frac{\kappa^2}{3}(\rho  _{+}-\rho _{-})\,,
\end{eqnarray}
and has the solution
\begin{eqnarray}
a(t)=\left[\frac{r}{2\lambda}\left(1+\frac{n^2}{6}\kappa^2\lambda t^2\right)\right]^{1/n}\,,
\end{eqnarray}
for $n\neq 0 $. Thus for any positive $n$ we get a regular bounce at $t=0$ with
minimal scale factor
\begin{eqnarray}
a_{0}=(r/2\lambda)^{1/n}\,,
\end{eqnarray}
 e.g. for
dust $(n=3)$, radiation $(n=4)$, stiff (holographic) matter ($n=6$) etc. Negative $n$ corresponds to matter
that violates the null energy condition (NEC). However, there is no big rip, but a bounce at the maximal
scale factor $a_{0}$. Such a phantom bounce was recently proposed in \cite{Brown:2004cs},
but no analytic solution was given. If we also introduce the characteristic time scale of the bounce
\begin{eqnarray}
t_0:=\sqrt{\frac{6}{\lambda}}\frac{1}{n\kappa}
\end{eqnarray}
we may write
\begin{eqnarray}
a(t)=a_0\left(1+\frac{t^2}{t_0^2}\right)^{1/n}\,. \label{aoft}
\end{eqnarray}

The relation between cosmic time $t$ and conformal time $\eta$ is given by
\begin{eqnarray}
\eta(t)&=&\int _0^t\frac{1}{a(x)}\,dx\\
&=&\frac{t}{a_0} F^{\frac{1}{2},\frac{1}{n}}_{\frac{3}{2}}\left(
-\frac{t^2}{t_0^2}\right)\,,
\end{eqnarray}
where $F^d_n(x)$ is a generalized hypergeometric function, but we will not make use of this relation in the following analysis.

Let us turn our attention to perturbations  and examine how they evolve through a big bang avoiding
bounce with minimal scale factor. This is the most interesting case \footnote{A big rip avoiding bounce is interesting on its own; see e.g. \cite{Nojiri:2004pf} for recent work in that direction. However, the need of having phantom matter and an additional time like dimension in a simple toy model seems to be, in our opinion, a little bit to much at this stage.}, because it could be seen as a simple toy model for a regularized version of the bounce occurring in the cyclic scenario. It is a toy model, since we remain at the four dimensional effective description throughout. We will comment on this issue in section \ref{interpret}. 

In the following, we have to be careful, since the so called ``sound speed"
$c_s^2=(\dot{p}_+-\dot{p}_-)/(\dot{\rho}_+-\dot{\rho}_-)$ diverges
near the bounce \footnote{This is a well known problem, see
e.g.\cite{Finelli:2003mc,Cartier:2004zn,Peter:2003}.}. To be
specific, for our background solutions one gets
\begin{eqnarray}
c_s^2=\frac{(a/a_0)^n(n/3-1)-(4n/3-2)}{(a/a_0)^n-2}\,,\label{soundvelocity}
\end{eqnarray}
so that $c_s^2$ diverges at $a^n=2^{1/n}a_0^n$, that is at $t=\pm t_0$. As a consequence, the usual equations
governing
the evolution of perturbations that involve $c_s^2$ can not be used near the bounce. Therefore, we will
derive the relevant equations of motion from first principles in the next section.

\section{Scalar perturbations \label{secpert}}
For the time being we will work with conformal time $\eta$.
The most general perturbed metric involving only scalar metric perturbations
in the longitudinal gauge is given by \cite{Bardeen:1980kt,Mukhanov:1990me,Brandenberger:2005be}
\begin{eqnarray}
d\,s^2&=&a^2\left[(1+2\Phi)d\,\eta ^2 -(1-2\Psi)\delta _{ij}d\,x^id\,x^j\right]\,.
\end{eqnarray}
Note that the two Bardeen potentials agree with the gauge invariant
scalar metric perturbations -- there is no residual
gauge freedom.

The perturbed energy momentum tensor is given by \cite{Mukhanov:1990me}
\begin{eqnarray}
( \delta T^{\mu}_{\,\,\nu})&=&
\left(
\begin{array}{cc}
-\delta\rho&(\rho _0+p_0)V_{,i}\\
-(\rho _0+p_0)V_{,i}&\delta p \delta ^i_j-\sigma _{;ij}
\end{array}
\right)\,,
\end{eqnarray}
where $\rho _0=\rho _+$ and $p _0=w\rho _0$ denote background quantities given by (\ref{rho+}) and $V$ is the
velocity potential so that
\begin{eqnarray}
(\delta u^\mu)=
\left(
\begin{array}{c}
0\\
\frac{V_{,i}}{a}
\end{array}
\right)\,.\label{defV}
\end{eqnarray}
Considering no anisotropic stress, that is $\sigma _{;ij}=0$, the off-diagonal Einstein equations yield
\begin{eqnarray}
\Phi=\Psi\,,
\end{eqnarray}
so that only one scalar metric degree of freedom is left.
Considering $\delta p=w\delta\rho+\tau\delta S$, where $\delta S$ is the entropy perturbation, and defining
$\xi :=\delta \rho/ \rho _0$ as well as $s:=\tau \delta S/\rho _0$ we have
\begin{eqnarray}
( \delta T^{\mu}_{\,\,\nu})&=&\rho _0
\left(
\begin{array}{cc}
-\xi&(1+w)V_{,i}\\
-(1+w)V_{,i}&(w\xi+s) \delta ^i_j
\end{array}
\right)
\end{eqnarray}
and from perturbing (\ref{smunu}) we get
 \begin{eqnarray}
( \delta S^{\mu}_{\,\,\nu})&=&\frac{\rho _0^2}{6}
\left(
\begin{array}{cc}
-\xi&(1+w)V_{,i}\\
-(1+w)V_{,i}&[(1+2w)\xi+s] \delta ^i_j
\end{array}
\right)\,.
\end{eqnarray}
With $\mathcal{H}=a^{\prime}/a$ the perturbed Einstein equations read
\begin{eqnarray}
&&\nabla ^2\Phi-3\mathcal{H}(\mathcal{H}\Phi +\Phi ^{\prime})=\frac{a^2}{2}\kappa^2\rho _0\xi\left(1-\frac{\rho _0}{\lambda}
\right)\,,\label{Einstein1}\\
 &&\Phi ^{\prime\prime}+3\mathcal{H}\Phi ^{\prime}+(2\mathcal{H}^{\prime}+\mathcal{H}^2)\Phi=
\frac{a^2}{2}\kappa^2\rho _0\label{Einstein2}\\
\nonumber&&\hspace{3cm}\times\left[w\xi+s-[(1+2w)\xi+s]\frac{\rho _0}{\lambda}\right]\,,\\
&&\left[\Phi\mathcal{H}+\Phi ^{\prime}\right]_{,i}=-\frac{a^2}{2}\kappa^2\rho_0V_{,i}(1+w)
\left(1-\frac{\rho _0}{\lambda}\right)\,.\label{Einstein3}
\end{eqnarray}

Note that the right hand side of (\ref{Einstein1}) will become zero
at $a^n=r/\lambda$. Henceforth, one might expect difficulties in
deriving a single second order differential equation for $\Phi$. A
method to deal with this problem, in the case of adiabatic
perturbation or $s=0$, was advocated in \cite{Peter:2002cn} and
subsequently used e.g. in \cite{Peter:2003,Finelli:2003mc} and the
previous versions of the article: one splits $\Phi$ in two
components, each of which satisfies a regular second order
differential equation. However, this method is inconsistent with
other constraint equations, like the fluid conservation equations in
the case of true two fluid models, as was explicitly shown in
\cite{Geshnizjani:2005hc}, or in our model the fact that the density
of the second fluid is related to the first fluid. There, an
alternative method was introduced, which we shall employ in the
following.

Since the universe was radiation dominated before the period of recombination,
let us restrict ourselves to $n=4$ such that
\begin{eqnarray}
a(x)=a_0\left(1+x^2\right)^{1/4}\,,\label{specifica}
\end{eqnarray}
Where, $x:=\frac{t}{t_0}$ is the rescaled time. Now, let us combine
(\ref{Einstein1}) and (\ref{Einstein2}) to
\begin{eqnarray}
\nonumber 0&=&\left(w-(1+2w)\frac{r}{\lambda a^4}\right)\left[-k^2\Phi-3\mathcal{H}(\mathcal{H}\Phi+\Phi^\prime)\right]\\
&&-\left(1-\frac{r}{\lambda a^4}\right)\left[ \Phi^{\prime\prime}+3\mathcal{H}\Phi^{\prime}+(2\mathcal{H}^{\prime}+\mathcal{H}^2)\Phi\right]\,,
\end{eqnarray}
where we used $\rho_0=ra^{-4}$, since $n=4$ and $w=1/3$ for
radiation. Converting this equation to rescaled cosmic time $x$,
making use of the background solution (\ref{specifica}) for $a(x)$
and introducing the rescaled momentum
\begin{eqnarray}
\tilde{k}:=\frac{kt_0}{a_0}
\end{eqnarray}
yields
\begin{eqnarray}
\left(1-\frac{2}{1+x^2}\right)\ddot{\Phi}_k+\frac{1}{2}\frac{x}{1+x^2}\left(5-\frac{18}{1+x^2}\right)\dot{\Phi}_k&&\label{maineom}\\
\nonumber -\left(\frac{1}{(1+x^2)^2}-\left[\frac{1}{3}-\frac{10}{3}\frac{1}{1+x^2}\right]\frac{\tilde{k}^2}{\sqrt{1+x^2}}\right)\Phi_k&=&0\,,
\end{eqnarray}
where $\Phi_k$ is a Fourier-mode of $\Phi$ and \textit{dot} denotes
a derivative with respect to $x$.

Before we go ahead and solve this equation in different regimes, let
us have a look at the problematic region around $x=\pm 1$, the
boundaries of the region where the \textit{Null Energy Condition} is
violated. The nontrivial question is whether $\Phi_k$ remains
continuous and smooth near these points or not. This question has been the
object of study in \cite{Geshnizjani:2005hc}, which provides
analytic solutions and a detailed discussion in the case of adiabatic perturbations.
Consequently, we will expand $\Phi_k$ near these points to find the
two independent analytic solutions, as proposed in
\cite{Geshnizjani:2005hc}.

\subsection{Approximate analytic solution}
Our final goal is to compute the spectrum
\begin{eqnarray}
\mathcal{P}_k:= k^{3}\left|\Phi_k^2\right|\propto k^{n_s-1}\label{defspectrum}
\end{eqnarray}
long after the bounce ($x\gg1$), while specifying the initial
conditions for each Fourier mode $\Phi_k$ well before the bounce
($x\ll-1$). To achieve that, we will find approximate analytic
solutions within different regions and match them together. In this
way, we will be able to compute analytically the spectral index
$n_s$.

Let us start in the limit $x\rightarrow-\infty$.
In that case one can introduce the gauge invariant variable $v_k$ in terms of which the action takes
the simple form of a scalar field (see \cite{Mukhanov:1990me} for a detailed discussion) so that the equation of motion reads
\begin{eqnarray}
v_k^{\prime\prime}+\left(c_s^2k^2-\frac{z^{\prime\prime}}{z}\right)v_k=0\,, \label{eomv}
\end{eqnarray}
with $c_s^2=1/3$ and
\begin{eqnarray}
z&:=&\frac{a\sqrt{\beta}}{\mathcal{H}c_s}\,,\\
\beta&:=&\mathcal{H}^2-\mathcal{H}^{\prime}\,.
\end{eqnarray}
Since the universe is radiation dominated, $z^{\prime\prime}$ vanishes and we can solve (\ref{eomv})
\begin{eqnarray}
v_k=\frac{3^{1/4}e^{-ik\eta/\sqrt{3}}}{\sqrt{k}}\,,
\end{eqnarray}
where we imposed quantum mechanical initial conditions for $v_k$, that is
\begin{eqnarray}
v_k(\eta_0)&=&k^{-1/2}M\,,\\
v_k^{\prime}(\eta_0)&=&ik^{1/2}N
\end{eqnarray}
and the normalization condition $NM^{*}+N^{*}M=2$.

The variable $v_k$ is related to $\Phi_k$ via
\begin{eqnarray}
\Phi_k=\sqrt{\frac{3}{2}}l_p\frac{\beta}{k^2\mathcal{H}c_s^2}\left(\frac{v_k}{z}\right)^\prime
\end{eqnarray}
in  that regime.
If we use the background solution in the limit $x\ll-1$ we get
\begin{eqnarray}
\nonumber \Phi_{k}&=&\alpha\frac{3^{3/4}}{2}\frac{1}{\tilde{k}^{3/2}x}\left(i-\frac{\sqrt{3}}{2}\frac{1}{\tilde{k}\sqrt{-x}}\right)
\exp\left(i\frac{2}{\sqrt{3}}\sqrt{-x}\tilde{k}\right)\\
&&\label{initialPhi}
\end{eqnarray}
as an approximate analytical solution for $\Phi$. The overall scale is given by
\begin{eqnarray}
\alpha:=\frac{l_p\sqrt{t_0}}{a_0^{3/2}}\,,
\end{eqnarray}
which is a free parameter in our model, subject to mild constraints, e.g. $t_0$ should be larger than the Planck time, but smaller than the time of nucleosynthesis. However, we will not need to specify its value if we scale the spectrum appropriately.

Similarly, for $x\gg1$ the general solution of
\begin{eqnarray}
\ddot{\Phi}_k+\frac{5}{2x}\dot{\Phi}_k+\frac{\tilde{k}^2}{3x}\Phi_k=0
\end{eqnarray}
can be written as
\begin{eqnarray} \label{postbouncesol}
\nonumber \Phi_k&=&\frac{C_1}{x}e^{i{2\over\sqrt{3}}\tilde{k}\sqrt{x}}(i-{\sqrt{3}\over 2}{1\over \tilde{k}\sqrt{x}})\\
&&+\frac{C_2}{x}e^{-i{2\over\sqrt{3}}\tilde{k}\sqrt{x}}(i+{\sqrt{3}\over
2}{1\over\tilde{k}\sqrt{x}})\,,\label{solaway}
\end{eqnarray}
where the $C_i$ are constants. For
\begin{eqnarray}
1\ll|x|&<&\sqrt{\left(\frac{3}{\tilde{k}^2}\right)^{2/3}-1}=:\tilde{x}\,,
\end{eqnarray}
 we can neglect the $\tilde{k}$-dependence of (\ref{maineom}), that is we can use
\begin{eqnarray} \label{afterNEC}
\left(1-\frac{2}{x^2}\right)\ddot{\Phi}_k+\frac{1}{2x}\left(5-\frac{18}{x^2}\right)\dot{\Phi}_k-\frac{\Phi_k}{x^4}=0\,,
\end{eqnarray}
which is solved by
\begin{eqnarray} \label{afternec}
\nonumber \Phi_k&=&C_3(x^2-2)^2x^{-(7-\sqrt{41})/4}F^{n_1}_{d_1}(x^2/2)\\
&&C_4(x^2-2)^2x^{-(7+\sqrt{41})/4}F^{n_2}_{d_2}(x^2/2)\,,\label{solbefore}
\end{eqnarray}
where
\begin{eqnarray}
n_1&=&\left[\frac{9+\sqrt{41}}{8},\frac{15+\sqrt{41}}{8}\right]\,,\\
d_1&=&1+\frac{\sqrt{41}}{4}\,,\\
n_2&=&\left[\frac{15-\sqrt{41}}{8},\frac{9-\sqrt{41}}{8}\right]\,,\\
d_2&=&1-\frac{\sqrt{41}}{4}\,,
\end{eqnarray}
and $F^n_d(z)$ is the generalized hypergeometric function.

Close to $x=\pm1$ we can Taylor-expand (\ref{maineom}) so that we get
\begin{eqnarray}
\nonumber \epsilon\ddot{\Phi}_k+\left(\frac{9}{4}\epsilon-1\right)\dot{\Phi}_k-\left(\frac{4-7\epsilon}{3\sqrt{2}}\tilde{k}^2+\frac{1}{4}-\frac{\epsilon}{2}\right)\Phi_k=0\,,&&\\
&&
\end{eqnarray}
where we introduced
\begin{eqnarray}
x=:\pm1 +\epsilon\,.
\end{eqnarray}
We will now follow \cite{Geshnizjani:2005hc} to solve this equation
near $x=\pm1$: if $\epsilon\ddot{\Phi}_k$ remains bounded, we get
one independent solution from which we can deduce the other one by
means of the Wronskian method. In this way, we arrive at
\begin{eqnarray}
\Phi_k=C_5\left(1-\left(\frac{1}{4}+\frac{4}{3\sqrt{2}}\tilde{k}^2\right)\epsilon\right)+C_6\epsilon^2\,.
\label{soltaylor}
\end{eqnarray}
Note that it would be justified to neglect the $\tilde{k}$
dependence in this solution. For the detailed reasoning as to why
this solution is appropriate we refer the reader to
\cite{Geshnizjani:2005hc}.

Last but not least, we have to find a solution for the actual bounce, that is for $|x|\ll1$ and $\tilde{k}\ll1$. The relevant equation of motion in this regime is
\begin{eqnarray}
\ddot{\Phi}_k+\frac{13}{2}x\dot{\Phi}_k+\Phi_k=0\,,\label{eomforsolbounce}
\end{eqnarray}
which is solved by
\begin{eqnarray}
\Phi_k=\frac{e^{-13x^2/8}}{\sqrt{x}}\left(C_7W^{m}_{n}(13x^2/7)+C_8M^{m}_{n}(13x^2/7)\right)\,,\label{solbounce}
\end{eqnarray}
with $m=-9/52$, $n=1/4$ and $W,M$ Whittakers functions.
Note that the solution is well behaved throughout the actual bounce.

All that is left to do is to match all the solutions smoothly. To be
specific one has to use (\ref{initialPhi}) for
$-\infty<x<-\tilde{x}$, (\ref{solbefore}) for
$-\tilde{x}<x<-1-\varepsilon$, (\ref{soltaylor}) for
$-1-\varepsilon<x<-1+\varepsilon$, (\ref{solbounce}) for
$-1+\varepsilon<x<1-\varepsilon$, (\ref{soltaylor}) for
$1-\varepsilon<x<1+\varepsilon$, (\ref{solbefore}) for
$1+\varepsilon<x<\tilde{x}$ and finally (\ref{solaway}) for
$\tilde{x}<x<1/(4\tilde{k})$ with some small $\varepsilon$. This
straightforward but tedious calculation results in a spectral index of
\begin{eqnarray}
n_s=-1\,, \label{analyticresult}
\end{eqnarray}
which is independent of the choice of $\varepsilon$ \footnote{Note that the amplitude depends on $\varepsilon$.
The reason the index is not effected by the choice of $\varepsilon$ lies in the fact that (\ref{soltaylor})
becomes independent of $\tilde{k}$ if  $\tilde{k}\ll1$.}.
This index is of course in conflict with the observed nearly scale invariant spectrum.

A few words regarding the validity of the matching procedure might
be in order. Lets first consider the matching at $x=\pm\tilde{x}$:
To either consider the term $\propto \tilde{k}^2$ or the term
stemming from $\dot{H}+2H^2$ in (\ref{maineom}) and match at the
point where both are equal is indeed a standard procedure that was
employed and tested (e.g. numerically) at various instances. For
example, in the framework of a bouncing universe it was used in
\cite{Peter:2002cn} and verified numerically in the limit
$\tilde{k}\ll1$, \footnote{Note that in \cite{Peter:2002cn}
additional matchings and approximations later on were needed, since
no full analytical solution was available.}. One can simply check
the validity of this matching for our case since the solutions of
(\ref{afternec}) as well as equations (\ref{initialPhi}) and
(\ref{postbouncesol}), around $|x| \sim \tilde{x}$ behave as
\begin{equation}
\Phi_k= A_i+{B_i\over x^{3/2}},
\end{equation}
where $A_i$'s and $B_i$'s are constant. This ensures a smooth
matching of these solutions at these points. Also, one could
successfully use the same matching procedure in the study of
fluctuations in an inflationary universes, even though a more
elegant procedure is available (see e.g. Chapter 13 of
\cite{Mukhanov:1990me} for a detailed discussion). The remaining
matchings occur close to $\pm1$ and do not influence the spectral
index as long as we are confident that the solutions are regular (as is the case for our scenario),
 since all equations
dictating the matching conditions are independent of $\tilde{k}$
in the limit of $\tilde{k}\ll1$.

As a consequence, we will be able to give a simplified analytical argument in the next section yielding the same spectral index. Thereafter, we compare our result with related studies in the literature like \cite{Bozza:2005wn,Bozza:2005xs}.

\subsection{Simplified matching procedure}
We would like to provide some simple insight into our result of
$n_s=-1$ from the previous section. As we mentioned earlier, it can
be shown that the solutions of the corresponding equations around
$|x|\sim\tilde{x}$ have the following asymptotic $x$ dependence
\begin{eqnarray}
\Phi_k= A_1+{B_1\over x^{3/2}}~~~~~x\rightarrow -\tilde{x}^-\,,\\
\Phi_k= A_2+{B_2\over x^{3/2}}~~~~~x\rightarrow -\tilde{x}^+\,,\\
\Phi_k= A_3+{B_3\over x^{3/2}}~~~~~x\rightarrow +\tilde{x}^-\,,\\
\Phi_k= A_4+{B_4\over x^{3/2}}~~~~~x\rightarrow +\tilde{x}^+\,.
\label{finalmode}
\end{eqnarray}
 This makes an analytic tracking of the calculation very
 easy: on the one hand, the smooth matching conditions are satisfied by requiring
 \begin{eqnarray}
 A_1&=&A_2,~~B_1=B_2\\
 A_3&=&A_4,~~B_3=B_4,
 \end{eqnarray}
 and on the other hand, the proper $\tilde{k}$ dependence of $A_1$ and $B_1$ can be enforced through
the the appropriate initial conditions leading to (\ref{initialPhi}). Taylor expanding
(\ref{initialPhi}) in the limit $\tilde{k}\ll 1$ results in
\begin{eqnarray}
\Phi_k(x)\sim{i\over k^{3/2}}({2\over3}k^2-{\sqrt{3}\over 2}{i\over
k (-x)^{3/2}})\,,
\end{eqnarray}
so that we have
\begin{eqnarray}
\label{A1B1} A_1&\propto& k^{1/2}\,,
 \\ B_1&\propto& k^{-5/2}\,.
\end{eqnarray}
Next, note that the transfer functions relating $A_2$ and $B_2$ to the
post-bounce coefficients $A_3$ and $B_3$ have to be independent
of $\tilde{k}$, since the equations governing the regime in between
$-\tilde{x}$ and $\tilde{x}$ are independent of $\tilde{k}$. In fact,
our equation have regular solutions
throughout this region so that it is reasonable to deduce
\begin{eqnarray}
A_3&=&\epsilon_1A_2+\epsilon_2B_2\,,\\
B_3&=&\iota_1A_2+\iota_2 B_2\,,
\end{eqnarray}
where $\epsilon_1, \epsilon_2, \iota_1$ and $\iota_2$ have to be
some constant numbers. Consequently the total transfer functions
will be
\begin{eqnarray}
A_4&=&\epsilon_1A_1+\epsilon_2B_1\,,\\
B_4&=&\iota_1A_1+\iota B_1\,.
\end{eqnarray}
Finally, making use of the above equations and  $A_1\ll B_1$ for
small $\tilde{k}$ according to (\ref{A1B1}), we can calculate the
power spectrum for  $A_4$, the non-decaying mode of
(\ref{finalmode})
\begin{eqnarray}
\mathcal{P}_k:= k^{3}\left|A_4^2\right|\propto
k^{3}\left|B_1^2\right|\propto \tilde{k}^{-2}\,.
\end{eqnarray}
Henceforth, we conclude a spectral index of
\begin{equation}
n_s=-1\,.
\end{equation}

\subsection{Numerical treatment}
\begin{figure}[tb]
  \includegraphics[width=\columnwidth]{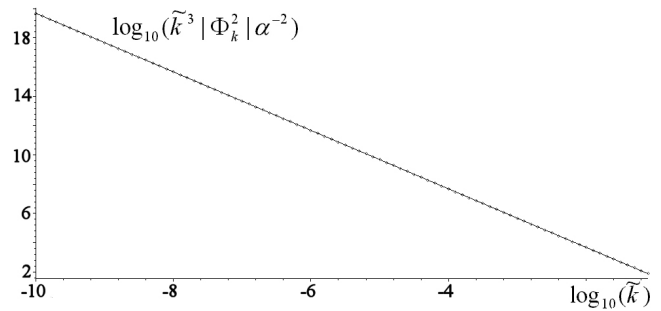}
   \caption{\label{Fig1} The numerical integration of (\ref{maineom}) yields the spectrum (\ref{defspectrum}) evaluated at $x=1/(4\tilde{k})$ (dots); the solid line corresponds to a spectral index of $n_s=-1$, in agreement with the analytic result (\ref{analyticresult}). Higher order effects in $\tilde{k}$ are accounted for by letting $\tilde{k}$ cover a broad range from $1$ down to $10^{-10}$.}
\end{figure}

In order to provide a quick check of our analytic arguments, we integrated equation (\ref{maineom}) with a 4/5th-order Runge-Kutta method, as implemented in MAPLE v9. Starting the numerical treatment at $-\tilde{x}$ with initial conditions given by (\ref{initialPhi}) and evaluating the final spectrum at $x=1/(4\tilde{k})$ yields  a spectral index of $n_s=-1$, Fig.\ref{Fig1}. The computations in MAPLE were done with 120 digits accuracy; nevertheless, it should be noted that the amplitude can not be recovered properly by the simple numerical method used, due to the non-trivial nature of the equation of motion near the bounce. However, this does not effect the spectral index since the actual bounce is $k$-independent, as can be seen in (\ref{eomforsolbounce}).

\subsection{Interpretation \label{interpret}}
Our result of the previous section clearly shows that for the
specific bouncing cosmology considered, both modes (the initial
constant one and the growing one) will pass through the bounce, but
it is the mode with the redder spectrum, i.e. the growing mode for
the Bunch-Davis vacuum, that dictates the final shape of the
post-bounce spectrum. Henceforth, we observe mode mixing in a model with only 
one degree of freedom and no spatial curvature. One has to be more careful to draw such
conclusion for more general cases involving two different fluids
(constrained by an adiabatic condition in other backgrounds), since
the matching procedure at $\pm\tilde{x}$ is not that simple in
general. To be more specific, the feature
of the bounce which is crucial for determining the approximate
behavior of the out coming spectrum, besides knowing the existence
and regularity of the solutions throughout the bounce region, is the
correction to the equation of motion (\ref{afterNEC}) due to the
second fluid. This is  the region where modes just turned non
oscillatory and corrections due to the bounce have to be considered,
because they are crucial for the solution.

Recently, a wide class of bouncing cosmologies with two independent
fluids was discussed in \cite{Bozza:2005wn,Bozza:2005xs} by Bozza
and Veneziano \footnote{See also \cite{Bozza:2005qg} for the most recent study}.
One might think that, in their notation, our class of
bouncing universes corresponds to $\alpha=0$, $c_a^2=(n-3)/3$,
$c_b^2=2n/3-1=2c_a^2+1$ and $n=4$. This would put it into region C
of \cite{Bozza:2005wn} and a spectral index of
\begin{eqnarray}
n_s&=&\frac{4(n-3)}{n-1}+1\\
&=&3
\end{eqnarray}
would result according to \cite{Bozza:2005wn}. However the model of
\cite{Bozza:2005wn} differs from our bouncing universe, since
$\alpha=0$ only enforces the adiabatic condition on each fluid
separately and while no intrinsic entropy modes for each fluid are
considered, two independent initial conditions for each of these
fluids are assumed. In our model the "two fluids" are indeed
related, so that the total pressure becomes a function of the total
energy (as a consequence only one initial condition can be
specified). In other words, we have enforced the adiabatic condition
in its strong interpretation. It should also be noted, the
approximation scheme employed there, differs considerably from ours,
involving an expansion in $k$ and a dimensional argument for the
contribution of the bounce (no explicit solution for the bounce was
used). Henceforth, the results of \cite{Bozza:2005wn,Bozza:2005xs}
can not be applied to our model.

A few words regarding Ekpyrotic/Cyclic scenarios are needed: in
\cite{Khoury:2001wf,Khoury:2001bz,Steinhardt:2001st} there is a
mechanism present that generates a scale invariant spectrum before
the bounce \cite{Gratton:2003pe}.
 It was shown in \cite{Creminelli:2004jg}, that this spectrum will not get transferred to the post-bounce era,
 if perturbations evolve independently of the details of the bouncing phase. Furthermore, the main conclusion of
  \cite{Bozza:2005wn,Bozza:2005xs} was, that a smooth bounce in four dimensions is not able to generate a scale invariant spectrum
  via the mode mixing technique. However, all these arguments were specific to bouncing
cosmologies in four dimensional gravity. So far, the effective four dimensional description
breaks down for all proposed cyclic models \cite{Tolley:2003nx} near the bounce, and the bounce itself
 is singular \cite{Turok:2004gb,Battefeld:2004mn}. Thus, it is of prime interest to search for a fully
 regularized bounce in cyclic/ekpyrotic models of the universe, so that one can compute the spectrum of
 perturbations explicitly \footnote{In \cite{Tsujikawa:2002qc} the bounce was regularized by higher-order
 terms stemming from quantum corrections, resulting in a spectrum that is not scale invariant; the model
 remained at the level of a four dimensional effective field theory. More recently, Giovannini followed this line of thought and
discussed perturbations in a five dimensional, regular toy model
\cite{Giovannini:2005fh}. One of his main conclusions was that the
new degrees of freedom associated with the extra dimension can be
interpreted as non-adiabatic pressure density
 variations from the four dimensional point of view.} in a full five dimensional setting
 - see e.g. \cite{Battefeld:2005wv} for a recent proposal and \cite{McFadden:2005mq} for an argument that mode mixing can occur in the full five dimensional setup. Our results confirm that it is challenging for
 such a model to succeed, but not impossible.

\section{Vector perturbations \label{secvector}} Vector
perturbations (VP) have recently caught attention in the context of
bouncing cosmologies, due to their growing nature in a contracting
universe \cite{Battefeld:2004cd,Giovannini:2004mc}. In the context
of a regularized bounce that we are discussing, VP remain finite and
well behaved, as we shall see now.

The most general perturbed metric including only VP is given by \cite{Bardeen:1980kt,Mukhanov:1990me}
\begin{eqnarray}
(\delta g_{\mu\nu})&=&-a^2
\left(\begin{array}{cc}
0&-S^i\\
-S^i&F^i_{\,\,,j}+F^j_{\,\,,i}
\end{array}\right)\,,
\end{eqnarray}
where the vectors $S$ and $F$ are divergenceless, that is $S^{i}_{\,\,,i}=0$ and $F^{i}_{\,\,,i}=0$.

A gauge invariant VP can be defined as \cite{Battefeld:2004cd}
\begin{eqnarray}
\sigma ^i&=&S^i +F^{i\,\prime} \,.
\end{eqnarray}

The most general perturbation of the energy momentum
tensor including only VP is given by \cite{Bardeen:1980kt}
\begin{eqnarray}
\nonumber (\delta T^{\mu}_{\,\,\nu})&=&
\left(
\begin{array}{cc}
0&(\rho_0 +p_0)V^i\\
-(\rho_0 +p_0)(V^i+S^i)&-p_0(\pi ^i_{\,\,,j}+\pi ^j_{\,\,,i})
\end{array}
\right)\,,\\
&& \label{VPemt}
\end{eqnarray}
where $\pi ^i$ and $V^i$ are divergenceless. Furthermore the perturbation in the 4-velocity is related to $V^i$ via
\begin{eqnarray}
(\delta u^\mu)=
\left(
\begin{array}{c}
0\\
\frac{V^i}{a}
\end{array}
\right)\,.
\end{eqnarray}
Gauge invariant quantities are given by
\begin{eqnarray}
\theta ^i&=&V^i-F^{i\,\prime}
\end{eqnarray}
and $\pi ^i$.

From now on we work in Newtonian gauge, where $F^i=0$ so that $\sigma$
coincides with $S$ and $\theta$ with $V$. Note that there is no residual
gauge freedom after going to Newtonian gauge. For simplicity we assume the absence of anisotropic stress, $\pi ^i=0$, so that
\begin{eqnarray}
\nonumber (\delta S^{\mu}_{\,\,\nu})&=&\frac{1}{6}\rho_0
\left(
\begin{array}{cc}
0&(\rho +p)V^i\\
-(\rho +p)(V^i+S^i)&0
\end{array}
\right)\,. \\
&&
\end{eqnarray}
The resulting equations of motion read
\begin{eqnarray}
\frac{1}{2a^2}\nabla^2S^i&=&-\kappa^2\rho_0(1+w)\left(1-\frac{\rho_0}{\lambda}\right)V^i\,,\label{eomV}\\
0&=&-\frac{1}{2a^4}\partial_\eta\left(a^2\left(S^j_{\,\,,i}+S^i_{\,\,,j}\right) \right)\,.\label{eomS}
\end{eqnarray}
Equation (\ref{eomS}) is easily intergrated for each Fourier mode, yielding
\begin{eqnarray}
S_{(k)}^i=\frac{C_{(k)}^i}{a^2}\,,
\end{eqnarray}
where $C_{(k)}^i$ is time independent. Since $a>0$, there is no divergent mode present. Equation (\ref{eomV}) is an algebraic equation for each Fourier mode of the velocity perturbation $V^i$.
Since $(1-\rho_0/\lambda)$ becomes zero shortly before and after the bounce, it seems like $V^i$ has to diverge before the bounce -- however, all that this is telling us is that $V^i$ is not the right variable to focus on: we should focus on the combination of $V^i$ and $\rho_0$ that appears in the perturbation of the effective total energy momentum tensor, that is $\rho_0(1+w)\left(1-\rho_0/\lambda\right)V^i$.
This combination is well behaved, and in fact one can check that it is always smaller than either $\rho_+-\rho_-$ or $p_+-p_-$ if we impose appropriately small initial conditions for $C_{(k)}^i$. Thus the perturbative treatment of VP is consistent.

\section{Tensor perturbations \label{sectensor}}

In this section we calculate the amplitude and spectral index of
gravity waves \footnote{We thank A. Starobinsky for pointing out pioniering work
on gravitational waves generated in a cosmological model
with a regular bounce \cite{Starobinsky:1979ty}.}, corresponding to tensor perturbations of the
metric, in a radiation dominated universe. Because the complete
five dimensional calculation is very complicated we will continue
using the theory of linear perturbations in a four dimensional
effective field theory. We include only the corrections $\propto
S^\mu_{\,\nu}$, due to fifth dimension, entering through the
evolution of the background.

In the linear theory of cosmological perturbations, tensor
perturbations can be added to the background metric via
\cite{Mukhanov:1990me}
\begin{equation}
ds^2=a^2[d\eta^2-(\delta_{ij}+h_{ij})dx^idx^j]\,,
\end{equation}
where $h_{ij}$ has to satisfy
\begin{equation}
h^i_i=h^{|j}_{ij}=0\,.
\end{equation}
Thus $h_{ij}$ has
only two degrees of freedom which correspond to the two polarizations
of gravity waves. One can show (see \cite{Mukhanov:1990me}) that
the amplitude  $h$ for each of these polarizations has to obey
\begin{equation}
\mu^{\prime\prime}-\nabla^2\mu-{a^{\prime\prime} \over a}\mu=0\,,
\end{equation}
where $\mu=ah$. Note that gravity waves do not couple to energy or
pressure and that they are gauge invariant from the start.

\begin{figure}[tb]
  \includegraphics[width=\columnwidth]{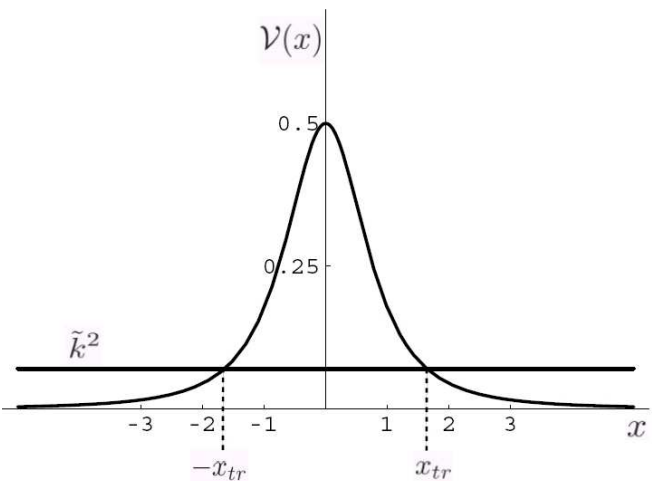}
   \caption{\label{V} ${\cal V}(x)$ sets a scale for the behavior of
each mode, it rises up to ${1\over 2}$ around the bounce but falls
off very quickly as $x\rightarrow \pm \infty$. }
\end{figure}

If the second term dominates over ${\mu a^{\prime\prime}/a}$ in
the above equation $\mu$ will exhibit oscillatory solutions, and
if the gradient term is negligible we will have $\mu\sim a$,
or, in other words, $h$ is almost constant. Using the standard
Fourier decomposition we get for each Fourier mode
\begin{equation} \label{eqgrav}
\mu_k^{\prime\prime}+\left(k^2-{a^{\prime\prime} \over
a}\right)\mu_k=0\,.
\end{equation}
Therefore, ${a^{\prime\prime}/a}$ sets a scale for the behavior of
each mode. This is very similar to the role of the Hubble radius for
inflationary scenarios, but rather than being almost constant,
${a^{\prime\prime}/a}$ is more like a potential barrier in quantum
mechanics:
\begin{equation} \label{eqV}
{\cal V}(x):= \left({t_0 \over a_0}\right)^2{a^{\prime\prime}\over a}={1\over 2} {1\over (1+x^2)^{3/2}}\,.
\end{equation}
As Fig.~\ref{V} shows, ${\cal V}(x)$ rises up to ${1\over 2}$ around
the bounce but falls off very quickly as $x\rightarrow \pm
\infty$.

Thus, for $\tilde{k}^2\gg {1 \over 2}$, the solutions of
equation (\ref{eqgrav}) are oscillatory. This implies that we
can set our initial conditions according to the Bunch-Davis vacuum at the initial time
$\eta_i$. We get
\begin{equation}
\mu_k(\eta)={1 \over \sqrt{2k}}e^{-ik(\eta-\eta_i)}\,,
\end{equation}
long before the bounce and it will remain basically unchanged
throughout the bounce. We easily compute the spectrum of
perturbations and the spectral index for these modes to be
\begin{eqnarray}
{\mathcal P}_k&:=& {2k^{3}\over M_p^2 a^2}\left|\mu_k^2\right|=
{2k^2 \over M_p^2 a^2}\propto
k^{n_t}\, ,\\
n_t&=&2~.
\end{eqnarray}

implying that, for $\tilde{k}^2\gg {1 \over 2}$, we are simply
left with vacuum fluctuations with an ever decaying amplitude.

Let us now turn our attention to the more interesting case:
$\tilde{k}^2\ll {1 \over 2}$.
As Fig.~\ref{V} demonstrates for these modes, we have transitions
from the regime $\tilde{k}^2\gg {\cal V}(x)$ (i.e. $k^2\gg
{a^{\prime\prime}/a}$) to the regime $\tilde{k}^2\ll {\cal
V}(x)$ at the transition points $\pm x_{_{tr}}$($\pm\eta_{_{tr}}$).
Here $x_{_{tr}}$ ($\eta_{_{tr}}$) is taken to be positive and
using (\ref{eqV}) we get
\begin{equation}
x_{_{tr}}\simeq2^{-1/3}~\tilde{k}^{-2/3}\,.
\end{equation}
For $ |x| \gg x_{_{tr}} $ ($|\eta|\gg \eta_{_{tr}}$) we can
simply approximate the solutions of equation (\ref{eqgrav}) to be
oscillatory
\begin{eqnarray}
\mu_k(\eta)&=&{1 \over \sqrt{2k}}[A_1e^{-ik(\eta-\eta_i)}+B_1e^{ik(\eta-\eta_i)}\bigl ]\,, ~~~\eta \ll -\eta_{_{tr}} \nonumber \\ \label{solutionI}\\
\mu_k(\eta)&=&{1 \over \sqrt{2k}}\bigr
[A_3e^{-ik\eta}+B_3e^{ik\eta}\bigl ]\,, ~~~\eta \gg
\eta_{_{tr}} \label{solutionIII}
\end{eqnarray}
where $ A_1 $ and $B_1$ are constants set by initial
conditions. In the case of the Bunch-Davis vacuum we can set
$A_1=1$ and $B_1=0$. We will derive $A_3$ and $B_3$ by matching the solutions smoothly
at the transition points.

But first, we need
to estimate the solutions for $\tilde{k}^2\ll {\cal V}(x)$ (i.e.
$k^2\ll {a^{\prime\prime}/ a}$) that is for $|x|\ll x_{_{tr}}$.
Unfortunately, unlike the case of most potential barriers in
Quantum Mechanics, we can not use the WKB approximation to
calculate the outgoing spectrum, since the condition
\begin{eqnarray}
{\partial
(k^2-a^{\prime\prime}/a)/\partial \eta \over
(k^2-a^{\prime\prime}/a)^2}\ll 1
\end{eqnarray}
 is not satisfied in this regime.
Fortunately we can still approximate the solution of equation
(\ref{eqgrav}) perturbatevely in orders of $k^2$ \footnote{This
solution is constructed iteratively from the solutions of equation
(\ref{eqgrav}) for $k=0$. This approach is very helpful for
analyzing most models of bouncing universes, since the equations
governing the evolution of perturbations are similar during the bounce,
see for example \cite{Peter:2003}.}
\begin{eqnarray} \label{solutionII}
{\mu_k(\eta)\over a(\eta)}&=&A_2\left[1-k^2\int^\eta {d\tau \over
a^2}\int^\tau d\varsigma ~a^2\right]\\
\nonumber &&+B_2\int^\eta
{d\tau\over a^2}\left[1-k^2\int^\tau d\varsigma ~a^2\int^\varsigma
{d\rho \over a^2}\right] + \cdot\cdot\cdot\,.
\end{eqnarray}
For $|\eta| \approx \eta_{_{tr}}$, since
\begin{eqnarray}
\int_0^\infty {d\eta \over a^2}&=& 2.60{t_0 \over a_0^3}
\,,\\a&\approx&  a_0|x|^{1/2}\sim a_0\tilde{k}^{-1/3}
\end{eqnarray}
and also\begin{equation} k^2\int^\eta_0 {d\tau \over
a^2}\int^\tau_0 d\varsigma ~a^2\approx {2\over 3}\, \tilde{k}^2
|x| ~\sim {2\over 3}\tilde{k}^{4/3}\,,
\end{equation}
 one can further approximate the result of
 equation (\ref{solutionII}) to
\begin{equation} \label{apsolutionII}
\mu_k(x)=A_2a_0|x|^{1/2}+B_2{t_0\over a_0^2}{|x|\over
x}[2.6|x|^{1/2}-2]\,.
\end{equation}
Notice that $\mu_k^{\prime\prime}(\eta_{_{tr}})=0$ for the above solution,
since
\begin{equation}
 {d \mu_k \over d\eta}= {1\over 2}\left[ \pm{ A_2 a_0^2\over t_0}+2.60{ B_2\over
 a_0}\right]\sim\rm{Const.}
\end{equation}
at $|\eta|\approx\eta_{_{tr}}$. Therefore, $\mu_k$ can be smoothly matched to our solutions from
equations (\ref{solutionI}) and (\ref{solutionIII}) at the transition
points $\pm x_{_{tr}}$ by equating $\mu_k$ and $\mu_k^\prime$. We
obtain the transfer function by relating the coefficients $A_1$
and $B_1$ to $A_3$ and $B_3$. These satisfy the relations
\begin{eqnarray}
A_3&\simeq& e^{ik(2\eta_{_{tr}}+\eta_i)}\bigl [0.65~ i
\tilde{k}^{-1}+2.32 \tilde{k}^{-1/3}\bigr
] A_1  \label{transmatrix1}\\
&&+ e^{-ik\eta_i}[0.65~ i \tilde{k}^{-1}]B_1\,,
\nonumber\\
B_3&\simeq&e^{ik\eta_i}[-0.65~i\tilde{k}^{-1}]A_1  \label{transmatix2}\\
&&+e^{-ik(2\eta_{_{tr}}+\eta_i)}\bigl [-0.65~ i
\tilde{k}^{-1}+2.32 \tilde{k}^{-1/3}\bigr ] B_1\,.\nonumber
\end{eqnarray}
The above result is valid for general initial conditions. We first
note that, to lowest order in $k^{-1}$, the constant mode in
the pre-collapse phase (growing mode for h), matches onto the
growing mode in the post-collapse phase (constant mode for h),
without any change in the spectral shape. As we mentioned before,
for a Bunch-Davis vacuum one simply takes $A_1=1$ and $B_1=0$. The
resulting power-spectrum for such a case is
\begin{equation} {\cal P}_k=13.44\left({\tilde{M}_p \over
M_p}\right)^6 ~x^{-1} \sin^2\left(k(\eta-\eta_{_{tr}}) \right)\,.
\end{equation}
The above spectrum is oscillatory and, for $k(\eta-\eta_{_{tr}})\gg
1$, it is oscillating so fast that one could conclude an almost flat ($n_t=0$) effective spectrum, with a decaying
amplitude. However, if $k(\eta-\eta_{_{tr}})\ll 1$ or in other
words $ 1 \ll x \ll 1/(4\tilde{k}^2)$, which is the case for
super-Hubble wavelengths, the above equation simplifies to
\begin{equation} {\cal P}_k=53.76 \left({\tilde{M}_p \over
M_p}\right)^6  \tilde{k}^2 \,.
\end{equation}

Thus, we get a non-decaying mode which has a blue spectrum
($n_t=2$). Notice, the amplitude of the power spectrum was dictated
directly by the scales and details of the bounce.
This result
is in agreement with the conclusions of \cite{Martin:2001ue}, where a
scale factor somewhat similar to our background scale
factor was used.

\section{Conclusion}
We constructed a wide class of regular bouncing universes, with a smooth transition from contraction to expansion.
This was motivated by open questions regarding the evolution of perturbations during singular bouncing
cosmologies in string cosmology, occurring e.g. in the cyclic scenario. The bounce is caused by
the presence of an extra time-like dimension, which introduces corrections to the effective four
dimensional Einstein equations at large energy densities. If matter respecting the null energy
condition is present, a big bang avoiding bounce results, and for phantom matter a big rip avoiding
bounce emerges. We find analytic expressions for the scale factor in all cases.

Having the Cyclic scenario in mind, we focused on a specific big
bang avoiding bounce with radiation as the only matter component. We
then discussed linear scalar, vector and tensor perturbations.

For scalar perturbations we compute analytically a spectral index of
$n_s=-1$, thus ruling out such a model as a realistic candidate. We
also find the spectrum to be sensitive to the bounce region, in
agreement to the majority of regular toy models discussed in the
literature. We conclude that it is challenging, but not impossible,
for Cyclic/Ekpyrotic models to succeed, if one can find a
regularized version.

Next, we discussed vector perturbations which are known to be problematic in bouncing cosmologies,
 due to their growing nature. We checked explicitly that they remain perturbative and small if compared
  to the background energy density and pressure.

 We concluded with a discussion of gravitational waves.
 The transfer function was computed analytically and we showed that the shape of the spectrum
 is unchanged in general. In the special case
  of vacuum initial conditions, the spectral index in post bounce era for super-Hubble wavelengths is blue.
  However, the amplitude turned out to be sensitive to the details of our model.

To summarize, having an explicit realization of e.g. the Cyclic
model of the universe that features a regular bounce it might be
possible to produce a scale invariant spectrum. Henceforth, the
challenge for the Cyclic model is twofold:
\begin{enumerate}
\item To find a convincing mechanism that regularizes the bounce.
\item To deal with the perturbations in a full five dimensional setup and to take into
account the presence of an inhomogeneous bulk as inevitable by the presence of branes.
\end{enumerate}
Preliminary results show that, by incorporating ideas of string/brane gas cosmology
(see \cite{Brandenberger:2005fb,Brandenberger:2005nz,Battefeld:2005av} for recent reviews),
a regularized bounce is possible (in preparation).

 \begin{acknowledgments}
We would like to thank Robert Brandenberger for helpful comments
and discussions and Scott Watson as well as Niayesh Afshordi,
Diana Battefeld, Zahra Fakhraai, Andrei Linde and V. Bozza for comments on the draft.
We would also like to thank the referee of PRD for drawing our attention to a possible problem associated with the splitting method used in previous versions of this article.
T.B. would like to thank the physics department of McGill
University and G.G would like to thank the physics department at
Harvard University for hospitality and support.
T.B. has been supported
in part by the US Department of Energy under contract
DE-FG02-91ER40688, TASK A.

\end{acknowledgments}

\end{document}